\documentclass[aps,prd]{revtex4-1}
\usepackage{amsmath, amssymb, graphicx, color, tikz, hyperref}
\def\Ngrids{N_g}
\def\sumgrids{\sum_{\rm grids}}
\def\gridsize{{\Delta x}}
\def\aap{Astronomy \& Astrophysics}

\def\apjl{Astrophysical Journal Letters}
\def\mnras{MNRAS}
\def\jcap{JCAP}
\def\prd{Physical Review D}
\def\prl{Physical Review Letters}
\def\apj{Astrophysical Journal}

\def\figwidth{0.75\textwidth}
\def\halfwidth{0.48\textwidth}

\begin{document}

\title{Curvature perturbations from kinetic preheating after $\alpha$-attractor inflation}

\author{Zhiqi Huang}
\email{huangzhq25@mail.sysu.edu.cn}
\affiliation{School of Physics and Astronomy, Sun Yat-sen University, 2 Daxue Road, Zhuhai, 519082, China}
\author{Xichang Ouyang}
\affiliation{School of Physics and Astronomy, Sun Yat-sen University, 2 Daxue Road, Zhuhai, 519082, China}
\author{Yu Cui}
\affiliation{School of Physics and Astronomy, Sun Yat-sen University, 2 Daxue Road, Zhuhai, 519082, China}
\author{Jianqi Liu}
\affiliation{School of Physics and Astronomy, Sun Yat-sen University, 2 Daxue Road, Zhuhai, 519082, China}
\author{Yanhong Yao}
\affiliation{School of Physics and Astronomy, Sun Yat-sen University, 2 Daxue Road, Zhuhai, 519082, China}
\author{Zehong Qiu}
\affiliation{School of Physics and Astronomy, Sun Yat-sen University, 2 Daxue Road, Zhuhai, 519082, China}
\author{Guangyao Yu}
\affiliation{School of Physics and Astronomy, Sun Yat-sen University, 2 Daxue Road, Zhuhai, 519082, China}
\author{Lu Huang}
\affiliation{CAS Key Laboratory of Theoretical Physics, Institute of Theoretical Physics, Chinese Academy of Sciences (CAS), Beijing 100190, China}
\author{Zhuoyang Li}
\affiliation{Department of Astronomy, Tsinghua University, No. 30 Shuangqing Road Beĳing 100084, China} 
\author{Chi-Fong Wong}
\affiliation{Faculty of Data Science, City University of Macau, Macau SAR, China}

\begin{abstract}
  Preheating at the end of inflation is a violent nonlinear process that efficiently transfers the energy of the inflaton to a second field, the preheat field. When the preheat field is light during inflation and its background value modulates the preheating process, the superhorizon isocurvature perturbations of the preheat field may be converted to curvature perturbations that leave an imprint on the cosmic microwave background and the large-scale structure of the universe. We use high-precision lattice simulations to study kinetic preheating after $\alpha$-attractor inflation, a case where the effective mass of the preheat field is naturally suppressed during inflation. By comparing the expansion e-folds between different Hubble patches, we find that the conversion from isocurvature perturbations to curvature perturbations is very inefficient and can hardly be detected by cosmological observations.
\end{abstract}

\maketitle

\section{Introduction \label{sec:intro}}

Observations in the last decade have made unprecedented progress in testing the inflationary paradigm that was originally proposed to solve the flatness and superhorizon correlation problems of the big-bang cosmology~\cite{Starobinskii_1979_spe,Kazanas_1980_dyn,Guth_1981_inf,Linde_1982_ane}. The well constrained spectral tilt of the primordial scalar perturbations, the upperbound of the primordial tensor perturbations, the smallness of the scalar spectral running, together with the lack of evidence for isocurvature perturbations and primordial non-Gaussianities have ruled out many inflationary scenarios including the simple quadratic ($m^2\phi^2$) and quartic ($\lambda\phi^4$) models~\cite{Aghanim_2020_par,Akrami_2020_NG, Ade_2018_BK,Akrami_2020_inf}. The remaining inflationary landscape, however, is still very populated~\cite{Martin_2024_cos}. In particular, a class of well-motivated $\alpha$-attractor models~\cite{Kallosh_2013_uni,Kallosh_2013_sup,Kallosh_2014_uni, Carrasco_2015_hyp, Levy_2024_alp, Kallosh_2024_sl}, which lies on the observational ``sweet spot'' and makes falsifiable predictions for the next generation of cosmic microwave background (CMB) experiments~\cite{Ade_2019_SO,Abazajian_2022_S4,Allys_2023_LB, Calabrese_2017_PIXIE}, has been extensively studied in the recent literature.

A standard inflationary model is followed by a reheating phase during which the inflaton transfers its energy to the matter and radiation of which the current universe is made up of.  In many models, reheating begins with a violent ``preheating'' phase where the energy is quickly dumped to a second field, the preheat field, via parametric resonance or tachyonic instability~\cite{Dolgov_1990_pro,Traschen_1990_par, Kofman_1994_reh, Kofman_1997_tow, Garcia_1998_pre}. For $\alpha$-attractor models, preheating can be naturally realized by taking into account the extra degrees of freedom in the supergravity construction of the model~\cite{Krajewski_2019_on, Iarygina_2019_uni} or significant kinetic coupling that arises from, e.g., hypernatural inflation~\cite{Linde_2018_hyp, Adshead_2023_kin}. 

Preheating may lead to non-thermal phase transitions and topological defects~\cite{Tkachev_1996_pha, Kofman_1996_non}, novel mechanisms of baryogenesis~\cite{Kolb_1996_gra, Anderson_1996_pre}, and relic short-wavelength gravitational waves (GWs)~\cite{Khlebnikov_1997_rel,  Huang_2011_art, Li_2020_GW, Jin_2020_GW, Krajewski_2022_GW, Adshead_2024_GW} that may be detected with future CMB experiments\footnote{The short-wavelength gravitational waves from preheating may make a significant contribution to the effective number of relativistic species, which can be accurately measured by future CMB experiments.} or novel techniques beyond laser interferometers~\cite{Aggarwal_2021_GW, Ito_2024_GW}. For multi-field inflation where large-scale isocurvature scalar perturbations can be generated during inflation, preheating may convert the isocurvature perturbations to curvature perturbations and leave an imprint on the CMB and large-scale structure of the universe. The basic mechanism is that the equation of state of the multi-field system (EOSOMS) during preheating is modulated by the isocurvature perturbations, yielding different expansion rates in different Hubble patches and generating curvature perturbations in the view of $\delta N$ formula~\cite{Salopek_1990_non, Battefeld_2008_mod, Chambers_2008_lat,Bond_2009_non,Kohri_2010_ont,Enqvist_2013_mod,Mazumar_2016_con, Artigas_2024_ext}. In the simplest construction, the preheat field is light during inflation and its long-wavelength perturbations modulate the preheating process. In this case, if the modulation of EOSOMS is triggered by some narrow bands of the preheat field, the curvature peaks are sheet-like and intermittent in configuration space, offering possible explanations of rare objects in CMB and large-scale structures~\cite{Bond_2009_non, Cruz_2005_det, Huang_2019_hig, Wang_2024_qua, Peebles_2023_flat, Lopez_2024_ring}.

This picture of modulated preheating is similar to the ``curvaton'' scenario where curvature perturbations originate from the post-inflation decay of a ``curvaton'' field with isocurvature perturbations~\cite{Mollerach_1990_curvaton,Lyth_2003_curvaton}. While the decay of curvaton can be perturbative and analytically calculated, the conversion from isocurvature perturbations to curvature perturbations for modulated preheating models involves complicated nonlinear dynamics and can only be evaluated with numeric simulations. Moreover, for most preheating models the decay of inflaton does not complete at the stage of parametric resonance, after which the multi-field system continues to evolve slowly~\cite{Kofman_1994_reh}. Long-term high-precision simulations are required to track the slow evolution until the EOSOMS converges to a constant, if it does. In the quartic potential model studied in Ref.~\cite{Bond_2009_non}, the EOSOMS after parametric resonance converges to a constant $1/3$ in an oscillatory way. The differences in expansion e-folds of simulations, the so-called $\delta N$ in separate universes, can be accurately estimated by averaging many oscillations. For more general models where the quadratic terms in the potential do not vanish, the 
 slow evolution of the EOSOMS is more complicated and requires more advanced techniques to extract the $\delta N$ information.

For kinetic preheating after $\alpha$-attractor inflation, the effective mass of the preheat field is naturally suppressed. It is therefore important to track the conversion from isocurvature perturbations to curvature perturbations, which may imply observational effects either supporting or excluding the model. We use lattice simulations to track the expansion histories of Hubble patches with different initial conditions, and compute the curvature perturbations with the $\delta N$ formula. We develop a novel method to decompose the evolution of the expansion e-fold into an oscillatory and a smooth polynomial component, extrapolation of which allows to extract the $\delta N$ information without long-time simulation of the slow-evolution stage.

This paper is organized as follows. Section~\ref{sec:model} introduces the model and the equations of motion. Section~\ref{sec:sim} describes the simulations. Results are given in Section~\ref{sec:res}. Section~\ref{sec:con} discusses and concludes. Throughout the paper we work with natural units $c=\hbar=1$ and a spatially flat Friedmann-Robertson-Walker (FRW) background metric $\mathrm{d}s^2 = \mathrm{d}t^2 -a^2(t)\mathbf{dx}^2$, where $t$ is the cosmological time. We use an overdot to denote the derivative with respect to $t$. The Hubble parameter is denoted as $H \equiv \frac{\dot a}{a}$. The reduced Planck mass is defined as $M_p \equiv 1/\sqrt{8\pi G}$, where $G$ is Newton's gravitational constant.

\section{Model \label{sec:model}}

We consider a two-field theory with a kinetic coupling between an inflaton field $\phi$ and a preheat field $\chi$, described by the action~\cite{Braglia_2020_gen, Kallosh_2022_dil}
\begin{equation}
  S = \int \sqrt{-g}\,\mathrm{d}^4x\, \left[\frac{M_p^2}{2}R + \frac{1}{2}(\partial \phi)^2 + \frac{1}{2}W(\phi)(\partial\chi)^2-V(\phi, \chi)\right].
\end{equation}
where $\sqrt{-g}\mathrm{d}^4x$ is the volume element and $R$ is the Ricci scalar. 

We employ lattice simulations of the fields $\phi$ and $\chi$ in an expanding FRW universe. The box size is taken to be on the order of horizon size, namely $H^{-1}$ at the end of inflation. Because the modulation of EOSMOS is dominated by the long-wavelength modes encoded in the generalized $\delta N$ formula, we may ignore sub-horizon metric perturbations~\cite{Bond_2009_non, Kohri_2010_ont, Huang_2019_hig}.   

For the lattice simulation, we use a program time $d\tau =  \frac{dt}{a^\beta}$, where $\beta$ is a freely chosen constant. For $\beta=0$ and $\beta=1$, the program time coincides with the cosmological time and the conformal time, respectively. A larger $\beta$ corresponds to smaller time steps and better accuracy. The choice of $\beta$ is therefore a balance between accuracy and speed. For a given choice of $\beta$, we use $y = a^{\frac{3-\beta}{2}}$ as a ``generalized scale factor''. On a cubic lattice with $N_g = n^3$ grids, comoving boxsize $L$, and comoving grid size $\gridsize = \frac{L}{n}$,  the discretized Lagrangian is
\begin{equation}
  \mathcal{L} = -\frac{12\Ngrids M_p^2{y^\prime}^2}{(3-\beta)^2} + \sumgrids \left\{\frac{y^2\left({\phi^\prime}^2 + W(\phi){\chi^\prime}^2 \right)}{2}
   - \frac{y^{\frac{2(1+\beta)}{3-\beta}}\left[ (\nabla\phi)^2 + (\nabla\chi)^2W(\phi) \right]}{2} 
    - y^{\frac{2(3+\beta)}{3-\beta}} V(\phi, \chi)\right\}, \label{eq:L}
\end{equation}
where a prime denotes derivative with respect to the program time $\tau$. The gradient term $(\nabla\phi)^2$ on the grid with indices $(i, j, k)$ is defined as 
\begin{eqnarray}
  \left[(\nabla \phi)^2\right]_{i, j, k} &\equiv &  \frac{1}{2\gridsize^2}\left[ \left(\phi_{i+1, j, k}-\phi_{i, j, k}\right)^2 + \left(\phi_{i, j, k}-\phi_{i-1, j, k}\right)^2  + \left(\phi_{i, j+1, k}-\phi_{i, j, k}\right)^2  \right. \nonumber \\
    && \left.+ \left(\phi_{i, j, k}-\phi_{i, j-1, k}\right)^2 + \left(\phi_{i, j, k+1}-\phi_{i, j, k}\right)^2 + \left(\phi_{i, j, k}-\phi_{i, j, k-1}\right)^2\right].
\end{eqnarray}
and $(\nabla \chi)^2$ is defined similarly.

From \eqref{eq:L} we obtain the canonical momenta
\begin{eqnarray}
  \pi_y &=&   -\frac{24}{(3-\beta)^2}M_p^2\Ngrids y^\prime ,\\
  \pi_\phi &=& y^2\phi^\prime ,\\
  \pi_\chi &=& y^2W(\phi)\chi^\prime.
\end{eqnarray}
and the Hamiltonian
\begin{equation}
  \mathcal{H} = \mathcal{H}_1 + \mathcal{H}_2 + \mathcal{H}_3 + \mathcal{H}_4, \label{eq:Hamil}
\end{equation}
where
\begin{eqnarray}
  \mathcal{H}_1 &=& -\frac{(3-\beta)^2}{48 M_p^2\Ngrids} \pi_y^2, \label{eq:H1} \\
  \mathcal{H}_2 &=& \frac{1}{2y^2}  \sumgrids \pi_\phi^2, \label{eq:H2}  \\
  \mathcal{H}_3 &=& \frac{1}{2y^2}  \sumgrids  \frac{\pi_\chi^2}{W(\phi)},  \label{eq:H3}  \\
  \mathcal{H}_4 &=& \frac{y^{\frac{2(1+\beta)}{3-\beta}}}{2} \sumgrids \left[ (\nabla\phi)^2 + (\nabla\chi)^2W(\phi) \right] + y^{\frac{2(3+\beta)}{3-\beta}}\sumgrids V(\phi, \chi). \label{eq:H4}
\end{eqnarray}

To evolve the two-field system on the lattice, we define the Poisson bracket operators $\hat{D} = \{\cdot, \mathcal{H}\}$ and $\hat{D}_i = \{ \cdot, \mathcal{H}_i\}$ for $i=1,2,3,4$. It follows then $\hat{D} = \hat{D}_1+\hat{D}_2+\hat{D}_3+\hat{D}_4$, and the time evolution operator for a finite time interval $\Delta \tau$ is
\begin{equation}
  e^{\hat{D}\Delta \tau} = e^{(\hat{D}_1+\hat{D}_2+\hat{D}_3+\hat{D}_4)\Delta \tau}. \label{eq:D1234}
\end{equation}
Because we are tracking the expansion e-folds which is closely connected to the total energy in the lattice via the Friedmann equations, it is crucial to use a high-order symplectic integrator to ensure long-term and high-precision energy conservation. We use 4-th order symplectic integrator to evolve the system, where Eq.~\eqref{eq:D1234} is written as
\begin{eqnarray}
  e^{\hat{D}\Delta \tau} &=& e^{c_1\hat{D}_1\Delta \tau/2} e^{c_1\hat{D}_2\Delta \tau/2} e^{c_1\hat{D}_3\Delta \tau/2} e^{c_1\hat{D}_4\Delta \tau} e^{c_1\hat{D}_3\Delta \tau/2}  e^{c_1\hat{D}_2\Delta \tau/2}  e^{(c_0+c_1)\hat{D}_1\Delta \tau/2} \nonumber \\
 &&\times e^{c_0\hat{D}_2\Delta \tau/2} e^{c_0\hat{D}_3\Delta \tau/2} e^{c_0\hat{D}_4\Delta \tau} e^{c_0\hat{D}_3\Delta \tau} e^{c_0\hat{D}_2\Delta \tau} e^{(c_0+c_1)\hat{D}_1\Delta \tau/2} \nonumber \\
 &&\times e^{c_1\hat{D}_2\Delta \tau/2} e^{c_1\hat{D}_3\Delta \tau/2} e^{c_1\hat{D}_4\Delta \tau} e^{c_1\hat{D}_3\Delta \tau/2}  e^{c_1\hat{D}_2\Delta \tau/2}  e^{c_1\hat{D}_1\Delta \tau/2} + O(\Delta\tau^5), \label{eq:Dapprox}  
\end{eqnarray}
where
\begin{equation}
  c_1 = \frac{1}{2-2^{1/3}},\ c_0 = 1- 2c_1.
\end{equation}

The operators $e^{\hat{D_i}\Delta \tau}$ ($i=1,2,3,4$) on the right-hand side of Eq.~\eqref{eq:Dapprox} can be executed precisely on a computer with only double-precision round-off errors, that is, $\sim 10^{-14}$. Thus the dominant error comes from the $O(\Delta\tau^5)$ term which can be efficiently suppressed by taking a relatively small time step.

The program expression of operator $e^{\hat{D}_1\Delta \tau}$  is
\begin{equation}
  y \rightarrow  y - \Delta\tau\frac{(3-\beta)^2\pi_y}{24M_p^2\Ngrids}.
\end{equation}
The operator $e^{\hat{D}_2\Delta \tau}$ is
\begin{eqnarray}
  \phi &\rightarrow & \phi + \frac{\Delta \tau}{y^2} \pi_\phi, \label{eq:D2_phi} \\
  \pi_y &\rightarrow & \pi_y + \frac{\Delta \tau}{y^3}\sumgrids \pi_\phi^2 ,
\end{eqnarray}
where Eq.~\eqref{eq:D2_phi} is applied on each grid.
The operator $e^{\hat{D}_3\Delta \tau}$ is
\begin{eqnarray}
  \chi &\rightarrow & \chi + \frac{ \Delta \tau}{y^2}\frac{\pi_\chi}{W(\phi)}, \label{eq:D3_chi} \\
  \pi_\phi &\rightarrow & \pi_\phi + \frac{ \Delta \tau}{2y^2} \pi_\chi^2\frac{W^\prime(\phi)}{W^2(\phi)}, \label{eq:D3_piphi}\\
  \pi_y &\rightarrow & \pi_y + \frac{\Delta \tau}{y^3}\sumgrids \frac{\pi_\chi^2}{W(\phi)},
\end{eqnarray}
where Eqs.~(\ref{eq:D3_chi}-\ref{eq:D3_piphi}) are applied on each grid. Finally, the operator $e^{\hat{D}_4\Delta \tau}$ is
\begin{eqnarray}
  \pi_\phi &\rightarrow & \pi_\phi + \Delta \tau\left[y^{\frac{2(1+\beta)}{3-\beta}}\nabla^2\phi - y^{\frac{2(3+\beta)}{3-\beta}}\frac{\partial V}{\partial \phi}- \frac{y^{\frac{2(1+\beta)}{3-\beta}}}{2}(\nabla\chi)^2W^\prime(\phi) \right] , \label{eq:piphi} \\
  \pi_\chi &\rightarrow & \pi_\chi + \Delta \tau\left[y^{\frac{2(1+\beta)}{3-\beta}}W(\phi)\nabla^2\chi - y^{\frac{2(3+\beta)}{3-\beta}}\frac{\partial V}{\partial \chi}\right] ,  \label{eq:pichi}  \\
 \pi_y &\rightarrow& \pi_y  - \Delta \tau\sumgrids \left\{\frac{1+\beta}{3-\beta}y^{\frac{3\beta-1}{3-\beta}}\left[(\nabla\phi)^2 + (\nabla\chi)^2W(\phi)\right] + \frac{2(3+\beta)y^{\frac{3(1+\beta)}{3-\beta}}}{3-\beta} V(\phi,\chi)\right\},
\end{eqnarray}
where Eqs.~(\ref{eq:piphi}-\ref{eq:pichi}) are applied on each grid. The Laplacian operator $\nabla^2\phi$  on the grid with indices $(i, j, k)$ is defined as 
\begin{equation}
  (\nabla^2\phi)_{i, j, k} = \frac{1}{\gridsize^2}\left(\phi_{i+1,j, k} + \phi_{i-1,j,k} + \phi_{i, j+1, k} + \phi_{i,j-1, k}+\phi_{i, j, k+1} + \phi_{i, j, k-1} - 6\phi_{i, j, k} \right),
\end{equation}
and $W(\phi)\nabla^2\chi$ is defined as
\begin{eqnarray}
  \left[W(\phi)\nabla^2\chi\right]_{i,j,k} &=& \frac{1}{2\gridsize^2}\left\{(\chi_{i+1,j, k}-\chi_{i, j, k})\left[W(\phi_{i+1, j, k})+ W(\phi_{i, j, k})\right]  \right. \nonumber\\
  &&  + (\chi_{i-1,j, k}-\chi_{i, j, k})\left[W(\phi_{i-1, j, k})+ W(\phi_{i, j, k}) \right] \nonumber\\
  &&  + (\chi_{i,j+1, k}-\chi_{i, j, k})\left[W(\phi_{i, j+1, k})+ W(\phi_{i, j, k}) \right] \nonumber\\
  &&  + (\chi_{i,j-1, k}-\chi_{i, j, k})\left[W(\phi_{i, j-1, k})+ W(\phi_{i, j, k}) \right] \nonumber\\
  &&  + (\chi_{i,j, k+1}-\chi_{i, j, k})\left[W(\phi_{i, j, k+1})+ W(\phi_{i, j, k}) \right] \nonumber\\
  &&  \left. + (\chi_{i,j, k-1}-\chi_{i, j, k})\left[W(\phi_{i, j, k-1})+ W(\phi_{i, j, k}) \right]\right\}.
\end{eqnarray}

Following Refs.~\cite{Adshead_2023_kin, Adshead_2024_GW}, we consider a dilaton-axion model with the symmetric T-model $\alpha$-attractor inflaton potential~\cite{Kallosh_2013_sup} 
\begin{equation}
  V(\phi, \chi) = \frac{m_\phi^2\mu^2}{2}\tanh^2\frac{\phi}{\mu} + \frac{1}{2}m_\chi^2\chi^2, \label{eq:Vphi}
\end{equation}
and a kinetic coupling
\begin{equation}
  W(\phi) = e^{\frac{2\phi}{\mu}}. \label{eq:Wphi}
\end{equation}
When $\mu \lesssim M_p$ and $m_\chi \lesssim m_\phi$, the self-interaction of inflaton and the strong kinetic coupling lead to efficient preheating~\cite{Adshead_2023_kin, Adshead_2024_GW}. When $\mu$ and the number of e-folds of inflation are specified, the inflaton mass can be determined by utilizing the amplitude of primordial power spectrum $\approx 2.1\times 10^{-9}$ at the pivot $k_{\rm pivot} = 0.05\,\mathrm{Mpc}^{-1}$. Taking $N_e$ ($N_e\sim$ 50-60)  e-folds of inflation, the slow-roll approximation gives
\begin{equation}
  A_s \approx \frac{H^2}{8\pi M_p^2\epsilon}, \label{eq:As_slowroll}
\end{equation}
where $\epsilon = \frac{\dot\phi^2}{2H^2}$ is the slow-roll parameter. The right-hand side of Eq.~\eqref{eq:As_slowroll} is evaluated $N_e$ e-folds before the end of inflation, which is defined by $\ddot a = 0$ or equivalently $\epsilon = 1$.

\section{Simulations \label{sec:sim} }

We take $\mu = 0.04M_p$, $m_\phi =7.2\times 10^{-6}M_p$, $m_\chi = 0.3m_\phi$ and assume single-field inflation along the $\phi$ direction.  These assumptions combined with Eq.~\eqref{eq:As_slowroll} leads to about $\sim 50.3$ e-folds. We start lattice simulations at $0.3$ e-folds before the end of inflation, where $\epsilon \approx 0.01$, and set the initial boxsize to be $1/H$. The advantage of starting the simulations slightly earlier than the end of inflation is that the initial conditions are simpler when slow-roll conditions still approximately hold.

The initial fluctuations of $\phi$ and $\chi$ are set with Bunch-Davies initial conditions in a classical form~\cite{Felder_2008_lat, Adshead_2023_kin}. When $m_\chi \sqrt{W(\phi)} \lesssim H$ during inflation, as in the case we study here, the mean $\chi$ value in the Hubble-size simulation box is $\lesssim H/\sqrt{W(\phi)}$. We define a dimensionless parameter $\lambda$ as the ratio between the initial mean $\chi$ in the simulation box and the initial $H/\sqrt{W(\phi)}$. In other words, the initial mean $\chi$ in the simulation box is
\begin{equation}
 \bar{\chi}_{\rm ini} =  \lambda \left(He^{-\frac{\phi}{\mu}}\right)_{\rm ini}.
\end{equation}
The conversion from the isocurvature perturbations in $\chi$ to the curvature perturbations is then encoded in the dependence of the expansion e-folds, taken at a fixed energy density (uniform energy gauge), on the simulation parameter $\lambda$. For horizon-size separate universes, the differences in expansion e-fold do not rely on Fourier modes that are much smaller than the horizon size. Low-resolution simulations are then sufficient for the $\delta N$ calculation~\cite{Bond_2009_non}. We mainly run simulations with $50^3$ grids and cross check against simulations with $100^3$ grids. The program parameter $\beta$ is taken to be $0.3$.

\begin{figure}
  \centering
  \includegraphics[width=\figwidth]{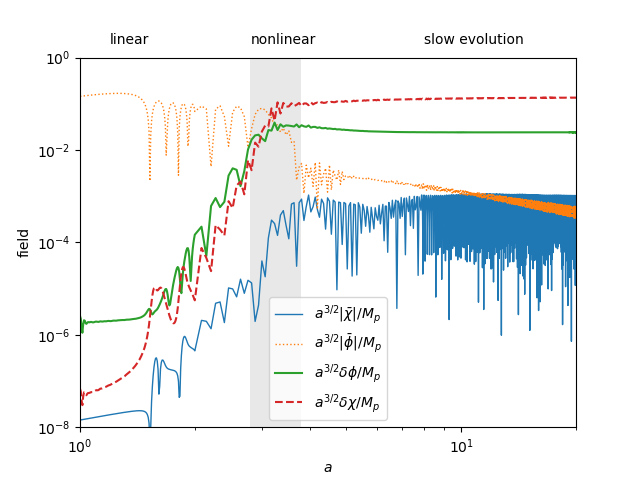}
  \caption{Evolution of field mean values $\bar{\phi}$ and $\bar{\chi}$ and root-mean-square fluctuations $\delta\phi$ and $\delta\chi$, all rescaled by $a^{3/2}$. \label{fig:fields}}
\end{figure}

Figure~\ref{fig:fields} shows the evolution of means,
\begin{equation}
  \bar{\phi} \equiv \frac{1}{N_g}\sum_{\rm grids} \phi; \  \  \bar{\chi} \equiv \frac{1}{N_g}\sum_{\rm grids} \chi; 
\end{equation}
and root-mean-square fluctuations
\begin{equation}
  \delta\phi \equiv \sqrt{ \frac{1}{N_g}\sum_{\rm grids} (\phi - \bar{\phi})^2}; \  \ \delta\chi \equiv \sqrt{ \frac{1}{N_g}\sum_{\rm grids} (\chi - \bar{\chi})^2},
\end{equation}
for a typical simulation. All amplitudes are rescaled by $a^{3/2}$ to approximately account for their damping due to Hubble friction. Parametric resonance starts with the linear growth stage where, despite the exponential growth of field fluctuations, gradient energy is still subdominant. After about one e-fold, a short period of nonlinear burst of field fluctuations (gray region) takes place, during which energy flows from the background inflaton oscillations to the gradient energy. After the stage of violent nonlinear growth of field fluctuations, the two-field system continues to evolve slowly. The remaining energy in the background inflaton oscillations decays slowly.

\begin{figure}
  \centering
  \includegraphics[width=\halfwidth]{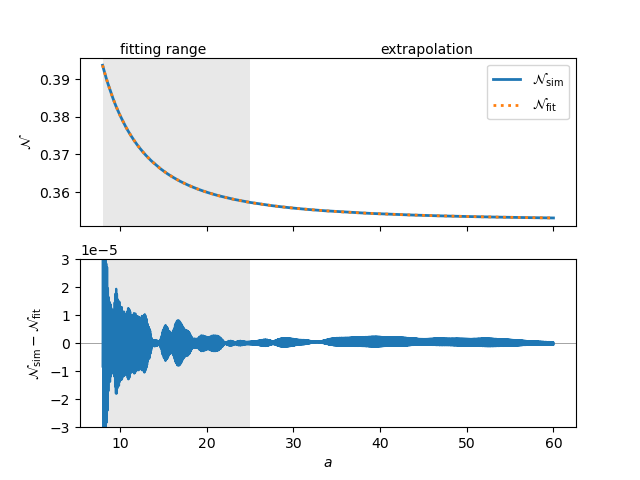}%
  \includegraphics[width=\halfwidth]{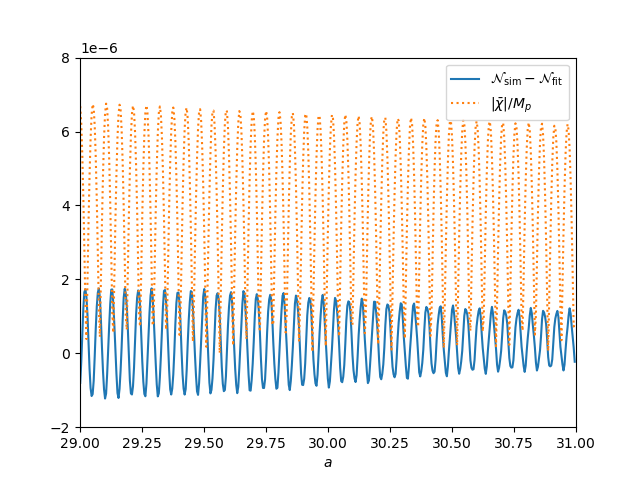}  
  \caption{Polynomial fitting of $\mathcal{N}_{\rm sim}$ (upper left panel), fitting residuals (lower left panel)  and the resonance between fitting residuals and the mean $\chi$ in the simulation box (right panel).\label{fig:Nfit}}
\end{figure}

In the slow-evolution stage, the two-field system is a mixture of harmonic oscillators with different effective masses. As the light modes redshift away, the EOSMOS gradually approaches $0$. In the $a\rightarrow \infty$ limit, $H^2a^3$ approaches a constant\footnote{This is an idealized picture that only includes the inflaton and the reheat field. In reality this process will be interrupted by subsequent reheating where the interaction with other fields cannot be ignored.}. Thus, we define an effective number of e-folds,
\begin{equation}
  \mathcal{N} \equiv \ln\left(\frac{H^{2/3} a}{H_{\rm ini}^{2/3}a_{\rm ini}}\right), \label{eq:Ndef}
\end{equation}
where $a_{\rm ini}\equiv 1$ and $H_{\rm ini}$ are scale factor and Hubble parameter at the beginning of simulation, respectively. The differences in $\ln a$ for simulations at a fixed energy density, or equivalently a fixed $H$, are identical to the differences in $\mathcal{N}$. The advantage of using $\mathcal{N}$ instead of $\ln a$ is that $\mathcal{N}$ approaches a constant in the slow-evolution stage and is therefore easier to extrapolate. The upper-left panel of Figure~\ref{fig:Nfit} shows the evolution of $\mathcal{N}$ from simulation\footnote{Figure~\ref{fig:Nfit} uses the same simulation that is used in Figure~\ref{fig:fields}.}, which we denotes  as $\mathcal{N}_{\rm sim}$. To compute the constant that $\mathcal{N}_{\rm sim}$ approaches in the $a\rightarrow \infty$ limit, we need to extract the smooth component of $\mathcal{N}_{\rm sim}(a)$. Because the energy density of massless modes decays as $a^{-4}$, the light modes and intermediate-mass modes may contribute $\sim a^{-n}$ ($n\leq 4$) terms. However, numeric experiments show that fitting $\mathcal{N}_{\rm sim}$ with a fifth-order polynomial of $\frac{1}{a}$
\begin{equation}
  \mathcal{N}_{\rm fit}(a) = c_0 + \frac{c_1}{a} + \frac{c_2}{a^2} + \frac{c_3}{a^3} + \frac{c_4}{a^4} + \frac{c_5}{a^5}, \label{eq:Nfit}
\end{equation}
works slightly better than fourth-order polynomial, in the sense that less simulation time is required to obtain an accurate $c_0$. The parameters $c_0, c_1,\ldots, c_5$ are determined by minimizing the integrated square difference
\begin{equation}
  E = \int_{a_{\rm fit, \min}}^{a_{\rm fit, \max}} \left[\mathcal{N}_{\rm sim}(a) - \mathcal{N}_{\rm fit}(a)\right]^2.
\end{equation}

Eq.~\eqref{eq:Nfit} is only a good approximation in the slow-evolution stage. For the model studied here, we should take $a_{\rm fit,\min} \gtrsim 5$ in the slow-evolution stage. Moreover, numerical experiments show that taking $a_{\rm fit,\max} = 25$ suffices to give a precision $\sim 10^{-6}$ of $c_0$, which is the target precision in the present work.

The lower-left panel of Figure~\ref{fig:Nfit} shows the difference $\mathcal{N}_{\rm sim} - \mathcal{N}_{\rm fit}$ for $a_{\rm fit, \min} = 8$ and $a_{\rm fit,\max} = 25$. Remarkably, the polynomial fit obtained at $a\in [8, 25]$ continues to work well at $a>25$. At $a>25$, the difference $\mathcal{N}_{\rm sim} - \mathcal{N}_{\rm fit}$ oscillates around zero with an amplitude $\lesssim 10^{-6}$. The right panel of Figure~\ref{fig:Nfit} zooms in some details of the oscillations of $\mathcal{N}_{\rm sim}-\mathcal{N}_{\rm fit}$. The oscillation period of $\mathcal{N}_{\rm sim}-\mathcal{N}_{\rm fit}$ is identical to that of $|\bar{\chi}|$, implying that the tiny oscillatory component in $\mathcal{N}_{\rm sim}$ is a physical component rather than numeric noises.

Hereafter for brevity, unless otherwise specified,  we implicitly refer $\mathcal{N}$ to  $\lim_{a\rightarrow \infty} \mathcal{N}=c_0$, which is practically obtained by running simulation till $a=25$ and fitting Eq.~\eqref{eq:Nfit}  between $a_{\rm fit,\min} = 8$ and $a_{\rm fit, \max}=25$. Because $\mathcal{N}$ relies not only on the initial mean $\chi$ (or $\lambda$), but also on the random seed to realize the Gaussian fluctuations of $\chi$ and $\phi$, we need to run many simulations for each $\lambda$ to average out the UV contribution.

\section{Results \label{sec:res}}

The simulation results are curvature perturbations on horizon scale during preheating, which is $\sim$ centimeter. To compare the theory with cosmological observations, we need to smooth the perturbations to $\gtrsim \mathrm{Mpc}$ scales. In a $\sim \mathrm{Mpc}$-size box, the initial $\sqrt{W(\bar{\phi})}\bar{\chi}$ in the Hubble patches during preheating, whose comoving size is $\sim e^{50}$ times smaller than $\mathrm{Mpc}$, follows a Gaussian distribution with standard deviation  $\sim \sqrt{50} \frac{H}{2\pi}$. This translates to a Gaussian smoothing window of $\lambda$ with standard deviation $\sigma_\lambda \approx \frac{\sqrt{50}}{2\pi}$. The smoothed $\mathcal{N}$ in $\mathrm{Mpc}$-size boxes is then defined as
\begin{equation}
  \mathcal{N}_{\rm Mpc}(\lambda) = \frac{1}{\sqrt{2\pi}\sigma_\lambda} \int_{-\infty}^\infty e^{-\frac{(\lambda^\prime - \lambda)^2}{2\sigma_\lambda^2}} \mathcal{N}(\lambda^\prime) d\lambda^\prime.
\end{equation}

\begin{figure}
  \centering
  \includegraphics[width=\figwidth]{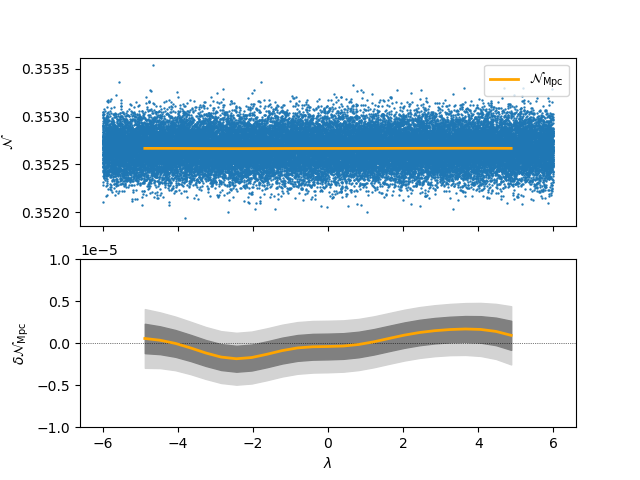}
  \caption{Upper panel: the raw results $\mathcal{N}$ (blue dots) and smoothed $\mathcal{N}_{\rm Mpc}$ (orange line); lower panel: the variation of $\mathcal{N}_{\rm Mpc}$ (orange line) and its 1$\sigma$ and 2$\sigma$ uncertainties (gray and light-gray areas). \label{fig:allN}}
\end{figure}

We run $38655$ simulations to study the dependence of $\mathcal{N}_{\rm Mpc}$ on the mean $\chi$ in $\mathrm{Mpc}$-size boxes. The raw results of $\mathcal{N}$ are shown as solid dots in the upper panel of Figure~\ref{fig:allN}. Despite the $\sim 10^{-3}$ scattering of $\mathcal{N}$, the smoothed $\mathcal{N}_{\rm Mpc}$ (orange line) is a constant within statistical errors ($\sim 10^{-6}$). The lower panel of Figure~\ref{fig:allN} zooms in the variation $\delta \mathcal{N}_{\rm Mpc}$ and its uncertainties, which indicate an upper bound $\lvert\delta \mathcal{N}_{\rm Mpc}\rvert\lesssim 10^{-6}$. This result is consistent with $\delta \mathcal{N}_{\rm Mpc} = 0$, i.e., no conversion from isocurvature fluctuations to curvature fluctuations on cosmological scales.

The uncertainties shown in the lower panel of Figure~\ref{fig:allN} only contain statistical errors. The four operators $\hat{D}_1, \hat{D}_2, \hat{D}_3, \hat{D}_4$ in the integrator~\eqref{eq:Dapprox} can be rearranged, for a total of 24 permutations. We use the 24 different integrators to confirm that the systematic error due to the choice of integrator is much smaller than the statistical errors. We also run 225 higher-resolution simulations with $100^3$ grids. The additional small-scale modes increase statistical errors, but the results are still consistent with $\delta\mathcal{N}_{\rm Mpc}=0$.

\section{Discussion and Conclusions \label{sec:con} }

The initial background value of the preheat field in principle can have an impact on the preheating process and modulate the expansion history. However, this phenomenon is hardly observed in numeric studies, because the chaotic non-linear dynamics of preheating often erases the memory of the initial conditions. There are exceptions, e.g., the modulated preheating model in Ref.~\cite{Bond_2009_non} which has a spindle-shaped potential well. The dynamics of the background trajectory of the inflaton field $\phi$ and the preheat field $\chi$ can be described as a billiard ball rolling back and forth in the spindle-shaped potential well. When the billiard enters one of the spindle arms, it can be locked there for a long time. This leads to an EOSOMS modulated by the isocurvature modes in the preheat field,  because the probability of entering the spindle arms is sensitive to the initial long-wavelength modes. The key ingredient here is the locking phenomenon that keeps the memory of initial conditions for a sufficiently long time.

The model studied here, kinetic preheating after $\alpha$-attractor inflation, has a novel feature that parametric resonance is mainly driven by a kinetic coupling term. However, the novel mechanism does not lead to a locking phenomenon that can keep the memory of initial conditions. A natural conjecture is then the conversion from the isocurvature perturbations to curvature perturbations is inefficient. The simulation results presented in this work are consistent with this conjecture. The upper bound $\lvert\delta\mathcal{N}_{\rm Mpc}\rvert \ll 10^{-6}$ implies that the curvature perturbations from preheating contribute no more than $\sim 0.1\%$ to the observed primordial scalar power spectrum $A_s \approx 2.1\times 10^{-9}$. Moreover, because the isocurvature perturbations in $\chi$ are uncorrelated with inflaton perturbations, three-point correlation vanishes to the lowest order. Thus, we conclude that kinetic preheating after $\alpha$-attractor inflation hardly produces any observable curvature perturbations.

\acknowledgments

This work is supported by National SKA Program of China No. 2020SKA0110402, the National Natural Science Foundation of China (NSFC) under Grant No. 12073088, and National key R\&D Program of China (Grant No. 2020YFC2201600). The computing resources are supported by Han Gao, Jianqiao Ge, Zhenjie Liu, Shitong Lu, Haitao Miao, Yan Su , Junchao Wang and Mingcheng Zhu via the MEET-U (``Make Everyone Explore The Universe'') program~\cite{MEETU-1}, where we make the simulation source code publicly available and allow astronomy amateurs to contribute their computing resources. 


%

\end{document}